\documentclass[a4paper,12pt,hidelinks]{article}
\usepackage[left=2.5cm,bottom=3cm,right=2.5cm,top=3cm]{geometry} 
\pdfoutput=1
\usepackage{graphicx}
\usepackage{xcolor}
\usepackage[pagebackref=true]{hyperref}
\usepackage{tikz} 
\usepackage{calrsfs}
\usepackage{cleveref}
\crefname{figure}{Figure}{Figures}
\usetikzlibrary{decorations.markings,decorations.pathmorphing}
\usetikzlibrary{intersections}
\usetikzlibrary{shapes.misc}
\tikzset{cross/.style={cross out, draw=black, minimum size=2*(#1-\pgflinewidth), inner sep=0pt, outer sep=0pt},
cross/.default={1pt}}
\tikzstyle{singularity}=[red!50!black,line width=0.6,decorate,
                         decoration={zigzag,amplitude=2,segment length=6.17}]
\colorlet{myred}{red!70!black}
\colorlet{mygreen}{green!70!black}
\colorlet{mydarkblue}{blue!50!black}
\colorlet{myblue}{blue!13!white!90!black}
\colorlet{myblue2}{blue!45!black!40!}
\colorlet{mygray}{gray!40!black}
\begin{document}
%
%
\begin{titlepage}
	\begin{flushright}
		IFT-UAM/CSIC-25-07
	\end{flushright}
	\vspace{.3in}
	\vspace{1cm}
	\begin{center}
		{\Large\bf\color{black} A Note on Black Hole Entropy and Wormhole  Instabilities}\\
		\bigskip\color{black}
		\vspace{1cm}{
			{\large J.~L.~F. Barb\'on$^a$ and E. Velasco-Aja$^{a,b}$}
			\vspace{0.3cm}
		} \\[7mm]
		$^a$	{\it {Instituto de F\'{i}sica Te\'orica IFT-UAM/CSIC, Universidad Aut\'onoma de Madrid, Cantobanco, 28049, Madrid, Spain}}\\[10pt]
		$^b$	{\it {Departamento de F\' \i sica Te\'orica, Universidad Aut\'onoma de Madrid, Cantoblanco 28049 Madrid, Spain}}\\[10pt]
		{\it E-mail:} \href{mailto:jose.barbon@csic.es}{\nolinkurl{jose.barbon@csic.es}}, \href{mailto:eduardo.velasco@uam.es}{\nolinkurl{eduardo.velasco@uam.es}}
	\end{center}
	\bigskip

\vspace{1cm}

\begin{abstract}
  We discuss recent approaches to the computation of black hole entropies through semiclassical estimates of appropriate state overlaps,  saturated by  Euclidean wormhole configurations. We notice that the relevant saddle-point manifolds may exhibit instabilities, thereby compromising the interpretation of the Euclidean path integral as a tool for computing positive-definite inner products.  We show that a proper treatment using a microcanonical formulation effectively addresses the puzzles posed by these instabilities. 
\end{abstract}
\end{titlepage}

\section{Introduction}
The Euclidean Gravitational Path Integral (GPI) formalism has a long history with many early successes \cite{Hawrev, GHP}, particularly in the context of semiclassical descriptions
of black hole thermodynamics \cite{GH}. Many of these early successes found a new life in the context of holography, including the reinterpretation of the `central mystery' posed
by the Gibbons--Hawking evaluation of black hole entropy from the saddle point value of the Euclidean gravitational action. The $A/4G$ microscopic degrees of freedom formally
accounted for by the classical saddle point would only show up with detail in the fully quantum,  dual boundary description. 

In this note, we discuss some aspects of recent proposals put forward in \cite{BLMS} (see also \cite{BLMSf, roberto, Leuven} for various generalizations of the same basic framework), which use Euclidean wormhole saddle points to estimate appropriately chosen state overlaps. This approach to
the calculation of black hole entropy emphasizes the fact that infinite families of  bulk states appear to be orthogonal to all orders in the effective field theory description in powers of  Newton's constant $G$, and
yet can have calculable universal overlaps of order ${\rm exp} (-c/G)$ which betray the finite-dimensionality of the black hole state space. 

The Euclidean wormhole contribution is interpreted as computing a somewhat `coarse-grained' version of the overlaps  (see \cite{Stanford},  \cite{Hartman} and \cite{Martinsolo} for relevant preliminary work). At a technical level, the analysis of \cite{BLMS} makes use of Lorentzian wormhole configurations stabilized by thin shells of dust, a choice with many advantages, such as a simple generalization to arbitrary dimensions and many types of black holes, a reasonably explicit holographic map to certain classes of microstates and a crucial `universality limit',  whereby the Newtonian mass of the dust shells is taken to be very large. This universality limit is used to take the last step and actually relating the microstate overlap estimates and the black hole entropy.

We observe that the relevant Euclidean wormhole saddle points can develop instabilities in this universality limit. These instabilities are associated to unstable thermodynamics and show up
as negative modes in one-loop fluctuation determinants \cite{gpy}. If these determinants are defined by analytic continuation, the resulting imaginary factors present a challenge to the interpretation of the GPI as a method to compute overlaps in a positive-definite inner product.  Our main conclusion is that a careful microcanonical treatment of the state overlaps solves all these issues in just the same way as they are solved in classic discussions of black hole thermodynamics using the Euclidean GPI. 

\section{Canonical Overlaps and their Instabilities}
\noindent
\subsubsection*{The Classical States}
\begin{figure}[htbp]
  \centering
    \begin{tikzpicture}[scale=2.5]
    \coordinate (-S) at (-1,-1); 
    \coordinate (-N) at (-1, 1);
    \coordinate (-E) at ( 0, 0);
    \coordinate (S) at ( 2,-1); 
    \coordinate (N) at ( 2, 1);
    \coordinate (E) at ( 3, 0);
    \coordinate (W) at ( 1, 0);
    \coordinate (c1) at ( 0.5, 1.02);
    \coordinate (c2) at ( 0.5, -1); 
     \draw[singularity] (-N) --  (N);
     \draw[singularity] (-S) --  (S);
     \draw[line width=1.75] (-N) --  (-S);
     \draw[line width=1.75] (N)  -- (S) ;
     \draw[line width=1.2,mydarkblue] (-N)-- (-E) -- (-S) ;
     \draw[line width=1.2,mydarkblue]  (-0.2,0.4) node[anchor=south ]{$\mathcal{H}(\mu_L)$} ;
     \draw[line width=1.2,mydarkblue]  (1.17,0.4) node[anchor=south ]{$\mathcal{H}(\mu_R)$} ;
     \draw[line width=1.2,mydarkblue] (N) -- (W)--(S);
     \draw[thick,myred, line width=2.25 pt] (c1)  -- (c2) ;
    \filldraw[thick,myred] (c1)  circle (0.2pt) ;
     \filldraw[thick,myred] (c2)  circle (0.2pt);
    \draw[myred]  (0.5,0) node[anchor=west]{$W_m$};
   \end{tikzpicture}
   \caption{\small The global structure of eternal AdS black holes with horizons ${\cal H} (\mu_L)$ and ${\cal H} (\mu_R)$ featuring long Lorentzian wormholes in their shared interior, supported by a shell of dust with  worldvolume $W_m$.} \label{fig:1}
  \end{figure}
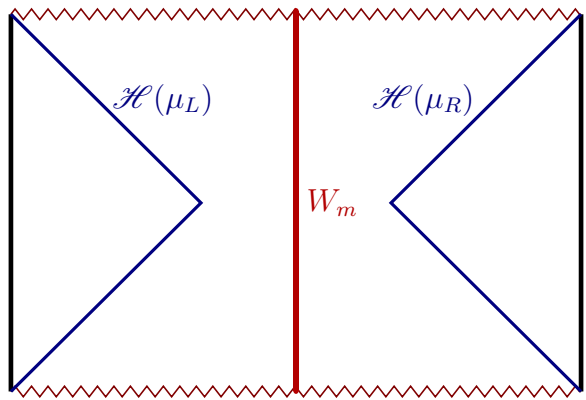

\noindent
We begin with a review of the basic setup explained in \cite{BLMS}. In \cref{fig:1} we show what is perhaps the simplest construction of  spacetimes which look like identical black holes from the outside but have vastly different interior geometries. Starting  from two eternal, two-sided AdS black holes, each having  a spherical thin shell of matter falling on one side, we cut out the asymptotic AdS regions lying above each shell and glue the two remaining spacetimes along the worldvolume of the shell. The result is a single two-sided AdS black hole with a large interior `wormhole' supported by the matter shell.  

The simplest matter model (and in many ways the more convenient one), is a fluid of dust with energy density $\sigma$. Einstein's equations in the thin-shell approximation imply that the Newtonian mass of the shell $m = \sigma(r) \,r^{d-1} \, {\rm Vol}({\bf S}^{d-1})$ is conserved along the free-fall trajectory $r(t)$ satisfying  
  \begin{equation}\label{weq}
  \left( {dr \over dt}\right)^2 + V_{m} (r) =0\,,\; \qquad V_{m} (r) = f(r) - \left({\Delta \mu \over m}-{4\pi G m \over (d-1) {\rm Vol}({\bf S}^{d-1}) r^{d-2}}\right)^2\;,
\end{equation}
where $f(r)$ is the red-shift function appearing in the AdS black hole metric with mass parameter $\mu$ and asymptotic radius of curvature $\ell$: 
\begin{equation}\label{adsbhl}
ds^2 = -f(r) \,dt^2 + {dr^2 \over f(r)} + r^2 \,d\Omega_{d-1}^2 \;, \qquad f(r) = 1+{r^2\over \ell^2} - {\mu \over r^{d-2}}\;,
\end{equation} 
and $\Delta \mu$ is given by  the mass parameter on the `side' of the spacetime from which (\ref{weq}) is being read, minus the mass parameter of the black hole across the shell.
With this convention for $\Delta \mu$, these equations parametrize the  worldvolume $W_m$ of the shell with mass $m$,  as seen from any of the two sides, left or right, with the understanding that the bulk geometries appearing in (\ref{adsbhl})   are specified independently on each side, with mass parameters $\mu_L$ and $\mu_R$. 

The constructed geometries are two-sided black holes with large, non-traversable wormholes in their shared interior.\footnote{One could consider left-right symmetric configurations and mod by the ${\bf Z}_2$ symmetry, to work with a `one-sided' version of the construction (see for instance \cite{hg}).} The wormhole's length is larger the `higher' the shells sit at their turning point ${\bar r}$, defined by the solution of  $V_{m} ({\bar r}) =0$.  We will be mostly interested in the regime of `high-flying' shells for which ${\bar r}$ is much larger than both the horizon radius $r_h$ and the AdS curvature radius $\ell$, scaling as ${\bar r} \sim (G\ell m )^{1 \over d-1}$, so that large Newtonian masses correspond to long wormholes. We conclude that the limit $m\to\infty$ yields a continuous infinity of classical spacetimes, all of them sharing the same exterior geometries but featuring interior Lorentzian wormholes of different lengths. We may consider a discrete family of ${\cal N}_w$ such solutions by picking Newtonian masses $m_k = k m$, labeled by the integer $k=1, \dots, {\cal N}_w$. In what follows, we will refer to the limit $m\to\infty$ as the `universality limit', in which all turning point radii ${\bar r}_k$ scale to infinity, thereby producing  ${\cal N}_w$ arbitrarily large, distinct Lorentzian wormholes, all of them inside black holes of fixed mass. 

\subsubsection*{The Quantum States}
\noindent
Let us now consider a set of ${\cal N}_w$ microscopic CFT states $| k \rangle$, which provide the non-perturbative version of the classical states just described. These states, having macroscopically different
internal geometries, appear to be orthogonal in the bulk description, and yet they must inhabit a subspace of the Hilbert space with dimension of order $\exp(S_L + S_R)$, with $S_{L,R}$ the black hole entropies of the left and right horizons. This implies that, when ${\cal N}_w \gg \exp(S_L + S_R)$, most of the $| k \rangle$ are linearly dependent. 

The authors of \cite{BLMS} characterize this situation in terms of the so-called Gram matrix, the ${\cal N}_w \times {\cal N}_w$ matrix of inner products ${\cal G}_{jk} = \langle j | k \rangle$, whose rank must be bounded by $\exp(S_L + S_R)$ no matter how large ${\cal N}_w$ is. The basic logic is that  suitable Euclidean wormhole geometries give semiclassical GPI approximations to the moments of the Gram matrix ${\rm Tr} \, {\cal G}^n$. From the knowledge of these moments, one can reconstruct the rank of ${\cal G}$ using resolvent techniques and confirm that ${\rm rank} ({\cal G}) \approx \exp(S_L + S_R)$ for asymptotically large ${\cal N}_w$.

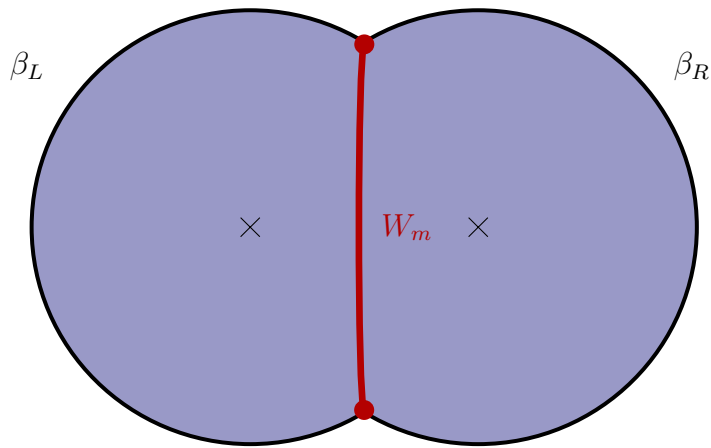
\begin{figure}[htbp]
 \centering
 \begin{tikzpicture}[scale=0.75]
 \def\outerRadius{3.8}
 \def\outerRadiusb{3.8}
 \def\yin{sqrt(5)}
 \coordinate (center1) at (-1,0);
 \coordinate (center2) at (3,0);
\pgfmathsetmacro\intersectX{\outerRadius*(sqrt(2)-1)}
\pgfmathsetmacro\intersectY{\outerRadius*(sqrt(2)-1)}
 \draw[name path=circle1,line width=3pt] (center1) circle (\outerRadius);
 \draw[name path=circle2,line width=3pt] (center2) circle (\outerRadiusb);
  \fill[myblue2] (center1) circle (\outerRadius) (center2) circle (\outerRadiusb);
 \path[name intersections={of=circle1 and circle2, by={A, B}}];
\draw[myred, thick, line width=2.5pt] (A) .. controls +(-0.125, -0.125) and +(-0.125, 0.125) .. (B);
 \fill[myred,thick] (A) circle[radius=5pt];
  \draw[myred,thick] (1.1,0) node[anchor=west] {$W_m$};
 \fill[myred,thick] (B) circle[radius=5pt];
 \draw (center2) node[cross=4pt] {};
 \draw (center1) node[cross=4pt] {};
 \filldraw [black] (-0.9*\outerRadius-1,0.75*\outerRadius) circle (0pt) node[anchor=east]{$\beta_L$};
 \filldraw [black] (0.8*\outerRadiusb+3.2,0.75*\outerRadiusb) circle (0pt) node[anchor=west]{$\beta_R$};
\end{tikzpicture}
\caption{\small The Euclidean manifold $X_{\mu_L, m, \mu_R}$ entering the GPI for the norm-squared of the PETS state. The crosses denote the left-right Euclidean horizons} 
 \label{fig:2}
 \end{figure}

 In order to infer the form of the CFT microstates, it is useful to switch to Euclidean signature. 
 In \cref{fig:2} we show the manifold  $X_{\mu_L, m, \mu_R}$, the Euclidean continuation of the two-sided black holes we started with, under a Wick rotation $t=-i\tau$. It consists of two Euclidean AdS black hole geometries $X_{\mu_{L}}, X_{\mu_R}$, cut through the dust worldvolume $W_m$ and glued back, side by side. Each of the Euclidean black hole solutions has a periodically identified time coordinate to avoid conical singularities: 
\begin{equation}\label{adsbh}
ds^2 (X_\mu) = f(r) \,d\tau^2 + {dr^2 \over f(r)} + r^2 \,d\Omega_{d-1}^2 \;, \qquad \tau \equiv \tau + \beta (\mu)\;.
\end{equation} 
 The function $\beta(\mu)$ is the inverse Hawking temperature of the AdS black hole with mass parameter $\mu$. It has a maximum ${\bar\beta}$ of $O(\ell)$,  it scales as $\beta(\mu) \sim r_h \sim \mu^{1\over d-2} $ for small $\mu$ and as $\beta(\mu) \sim \ell^2 /r_h \sim \ell^2 (\mu \ell^2)^{-1/d}$ for large $\mu$, where $r_h$ is the
 horizon radius. For each value of $\beta$ below the maximum, there are two possible manifolds $X_{\mu^\pm}$ corresponding to large black holes,   having $r_h > O(\ell)$, or small black holes, having $r_h<O(\ell)$. This description, particularly in what concerns the existence of the small black hole branch, requires $d>2$, a condition that we shall assume throughout this paper.

 According to the AdS/CFT rules, the asymptotic behavior of the Euclidean geometry suggests the form of the non-perturbative definition for the GPI on $X_{\mu_L, m, \mu_R}$. In the present case, we must have a generalized partition function
 \begin{equation}\label{genpart}
 Z_{\beta_L, {\cal O}^\dagger, \beta_R, {\cal O}} \equiv  {\rm Tr} \left({\cal O}^\dagger \;e^{-\beta_L H} \;{\cal O} \;e^{-\beta_R H}\right) \;,
\end{equation} 
where $H$ is the CFT Hamiltonian and ${\cal O}$ is a  CFT operator, acting non-locally over the spatial sphere of the CFT, and  representing the insertion of the dust shell  with mass $m$. Its detailed  UV structure is left unspecified as long as we remain within a low-energy, semiclassical description of the bulk. The relation between the CFT preparation inverse temperatures $\beta_{L,R}$ and the classical geometry is
\begin{equation}\label{sumrp}
\beta(\mu_s) = \beta_s + \Delta\tau_{W} \;,
\end{equation}
where $s= L, R$ is the `side' index, $\beta(\mu_s)$ is the identification interval for the periodic Euclidean time coordinate $\tau$ in each side,  and $\Delta\tau_W$ is the Euclidean time
extent of the dust worldvolume, as seen from any of the two sides,
\begin{equation}\label{dangle}
\Delta \tau_W = 2 \int_{\bar r}^\infty {dr \over f(r)} \sqrt{f(r) - V_{m} (r) \over V_{m} (r)}\;. 
\end{equation}

Cutting the CFT trace open in (\ref{genpart}) reveals that it can be interpreted as the squared norm, 
\begin{equation}\label{inpet}
\langle  \psi_{\beta_L, {\cal O}, \beta_R }\,|  \psi_{\beta_L, {\cal O}, \beta_R} \rangle =  Z_{\beta_L, {\cal O}^\dagger, \beta_R, {\cal O}}\; ,
\end{equation}
 of a so-called  PETS (Partially Entangled Thermal State), a 
generalization of the thermofield-double state in which the entanglement is disrupted by the insertion of an operator (see for example \cite{PETS}), 
\begin{equation}\label{PETO}
|\psi_{\beta_L, {\cal O},  \beta_R} \rangle = \sum_{n,m} \, |n^*\rangle_L\, \left(e^{-\beta_L H_L /2} \,{\cal O} \,e^{-\beta_R H_R /2}\right)_{n\,m}  \, | m \rangle_R \;,
\end{equation}
where $|n\rangle_{L,R}$ denote energy eigenstates of the (identical) `left' and `right' CFT Hamiltonians $H_L = H_R = H$, defined on $(d-1)$-dimensional spheres of radius $\ell$, and $|n^*\rangle$ stands for the CRT conjugate of the $|n\rangle$ state.  

The GPI approximation to (\ref{genpart}) thus defines a GPI version of the inner product
\begin{equation} \label{epi}
\overline{\langle  \psi_{\beta_L, {\cal O}, \beta_R }\,|  \psi_{\beta_L, {\cal O},  \beta_R} \rangle }= \overline{Z}[X_{\mu_L, m, \mu_R}] = \int_{X_{\mu_L, m,  \mu_R} } \,e^{-I[X_{\mu_L, m, \mu_R}]} \;. 
\end{equation}
Here and henceforth  an over-line denotes the bulk GPI evaluation of the quantity in question.  In this path integral, $I[X]$ is the Euclidean action of the bulk, minimally containing Einstein gravity, the dust fluid, and the relevant boundary terms required by the AdS/CFT rules.    As usual, these rules allow for non-perturbative corrections in the form of other saddle points with possibly different topology, such as the vacuum thermal AdS manifold, with topology ${\bf S}^1 \times {\bf R}^{d}$, which we denote $X_0$, and corresponds to the $\mu=0$ case of (\ref{adsbhl}), now admitting an arbitrary $\tau$ identification period, $\beta$. 

\subsubsection*{The Euclidean Wormholes}
\noindent
Let us now consider a cyclic product of microstate overlaps 
\begin{equation}\label{prod}
\langle \psi_{\beta_{L}, {\cal O}_1, \beta_{R }} | \psi_{\beta_{L } ,{\cal O}_2, \beta_{R } } \rangle \langle \psi_{\beta_{L}, {\cal O}_2, \beta_{R }} | \psi_{\beta_{L } ,{\cal O}_3, \beta_{R} } \rangle \cdots \langle \psi_{\beta_{L }, {\cal O}_n, \beta_{R }} | \psi_{\beta_{L }, {\cal O}_1, \beta_{R }} \rangle \;.  
\end{equation} 
Here, we arrange the same Euclidean preparation inverse temperatures, $\beta_{L, R}$, for all $n$ microstates participating in the cyclic product, in such a way that the microstates are only distinguished by the discrete choice of operators ${\cal O}_j$, with $j=1, \dots, {\cal N}_w$. Such cyclic products are building blocks for the moments of the Gram matrix ${\rm Tr} \,{\cal G}^n$. In the GPI construction, these states are specifically coarse-grained: roughly speaking, we forget all the information contained in the matrix elements $\langle m | {\cal O}_j | n\rangle$ except for the mass parameter $m_j$. If we imagine that this microscopic information is effectively washed away by some sort of statistical average, there will be a large  correlation between ${\cal O}_j$ and its Hermitian conjugate ${\cal O}_j^\dagger$. Therefore, one would expect that the operator coarse-graining would connect each `ket' (containing ${\cal O}$) with the consecutive `bra' (containing ${\cal O}^\dagger$) in  cyclic products  such as (\ref{prod}). An extreme version of this phenomenon can be readily seen by assuming that the operators ${\cal O}_j$ are drawn randomly with independent Gaussian statistics, i.e. we postulate a microscopic averaging rule:
\begin{equation}\label{microcg}
\overline{({\cal O}_j)_{l\,r}  ({\cal O}_k^\dagger)_{r' \,l'} }\; \Bigg |_{{\cal O}-{\rm averaged}}= {\rm C}_j \,\delta_{j\,k}\;\delta_{r\,r'} \,\delta_{l\,l'}\;,
\end{equation}
with ${\rm C}_j$ some operator-dependent constants (we also specify that the over-line here  refers to the operator averaging, rather than the one induced by the GPI). Performing this average in the cyclic product (\ref{prod}), using the explicit form of the states (\ref{PETO}) we find 
\begin{equation}\label{prodi}
\overline{\langle \psi_{i_1} | \psi_{i_2} \rangle \,\langle \psi_{i_2} | \psi_{i_3} \rangle \cdots \langle \psi_{i_n} | \psi_{i_1} \rangle }\; \Bigg |_{ {\cal O}-{\rm averaged}} = Z(n\beta_L) \,Z(n\beta_R)\, \prod_{j} {\rm C_j}\;.
\end{equation}
The operator averaging has taken the product of $n$ CFT traces into a single left-right product of traces. In other words, it decouples the left from the right but it couples the $n$
overlaps. This suggests that the GPI approximation to (\ref{prod}), 
 \begin{equation}\label{woh}
{\overline{\langle \psi_{i_1} | \psi_{i_2} \rangle \,\langle \psi_{i_2} | \psi_{i_3} \rangle \cdots \langle \psi_{i_n} | \psi_{i_1} \rangle} } = \overline{ Z}[X_{i_1 \dots i_n}] = \int_{X_{i_1, i_2, \dots i_n}} e^{-I[X_{i_1 \dots i_n}]}\;, 
\end{equation}
should involve connected manifolds $X_{i_1 \dots i_n}$, i.e. the $n$-replica wormholes. From the structure of (\ref{genpart}) and (\ref{inpet}), we deduce that they are  required to have the same conformal boundary as a disconnected product of $n$ CFT traces, that is to say, an $n$-fold product of ${\bf S}^1 \times {\bf S}^{d-1}$. Each ${\bf S}^1$ factor must have a pair of distinguished points, cutting the circle into two segments of length $\beta_L$ and $\beta_R$, corresponding to the endpoints of the shell's worldvolumes. The $n$ replicated ${\bf S}^1$ boundaries are cyclically ordered, in correspondence to the emission and absorption of the $n$ dust shells of masses $m_j$. 

\begin{figure}[htbp] 
\centering
\begin{tikzpicture}[scale=1.7]
    \def\a{1.15} 
    \def\b{.25} 
    \draw[line width=4 pt,rotate=45] (-2.3,-4.5) ellipse({\a} and {\b});
    \draw[line width=4 pt,rotate=-45] (2.3,-4.5) ellipse({\a} and {\b});  
    \filldraw[white] (-2.4,-2.45) .. controls +(0.4, -0.9)  and  +(-0.1, -0.27) ..(-2.4,-4)--(-.77,-5.64).. controls +(0.4, 0.5) and +(-0.4, 0.5) .. (0.77,-5.64)--(2.4,-4).. controls +(0.1, -0.27) and +(-0.4, -0.9) .. (2.4,-2.45) --(0.77,-.77).. controls +(-0.4, -0.5)  and +(0.4, -0.5) ..(-0.77,-0.77)-- (-2.4,-2.45)--cycle;
    \fill[myblue2] (-2.4,-2.45) .. controls +(0.4, -0.9)  and  +(-0.1, -0.27) ..(-2.4,-4)--(-.77,-5.64).. controls +(0.4, 0.5) and +(-0.4, 0.5) .. (0.77,-5.64)--(2.4,-4).. controls +(0.1, -0.27) and +(-0.4, -0.9) .. (2.4,-2.45) --(0.77,-.77).. controls +(-0.4, -0.5)  and +(0.4, -0.5) ..(-0.77,-0.77)-- (-2.4,-2.45)--cycle;
    \draw[line width=3 pt,rotate=45,dashed] (-2.3,-4.5) ellipse({\a} and {\b});
    \draw[line width=3 pt,rotate=-45,dashed] (2.3,-4.5) ellipse({\a} and {\b});   
    \fill[white,line width=2 pt,thick,rotate=-45] (2.3,-4.5) ellipse({\a} and {\b});
    \fill[myblue2,line width=2 pt,thick,rotate=-45] (2.3,-4.5) ellipse({\a} and {\b});
    \fill[white,line width=1 pt,thick,rotate=45] (-2.3,-4.5) ellipse({\a} and {\b});
    \fill[myblue2,line width=2 pt,thick,rotate=45] (-2.3,-4.5) ellipse({\a} and {\b});
    \filldraw[white,line width=2 pt,thick,rotate=45] (-2.3,0) ellipse({\a} and {\b});        
    \fill[myblue,line width=2 pt,thick,rotate=45] (-2.3,0) ellipse({\a} and {\b});
    \fill[white,line width=2 pt,thick,rotate=-45] (2.3,0) ellipse({\a} and {\b});
    \fill[myblue,line width=2 pt,thick,rotate=-45] (2.3,0) ellipse({\a} and {\b});
    \draw[rotate=45,line width=2 pt] (-2.3,0) ellipse({\a} and {\b});
    \draw[rotate=-45,line width=2 pt] (2.3,0) ellipse({\a} and {\b});
    \filldraw[mydarkblue] (-2.6,-3.2) circle[radius=0pt] node[]{$W_4$};
    \filldraw[myred] (2.65,-3.2) circle[radius=0pt] node[]{$W_2$};
    \filldraw[mygreen] (0,-0.6) circle[radius=0pt] node[]{$W_1$};
    \filldraw[orange] (0,-5.7) circle[radius=0pt] node[]{$W_3$};
    \draw[myred,thick, line width=2.5pt] (2.4,-4).. controls +(0.1, -0.27) and +(-0.4, -0.9) .. (2.4,-2.45); 
    \filldraw[myred] (2.4,-2.45) circle[radius=3pt] {};
    \filldraw[myred] (2.4,-4) circle[radius=3pt] {};
    \filldraw[mygreen] (-.77,-0.77) circle[radius=3pt] {};
    \filldraw[mygreen] (.77,-0.77) circle[radius=3pt] {};
    \draw[mygreen,thick, line width=2.5pt] (-0.77,-0.77) .. controls +(0.4, -0.5) and +(-0.4, -0.5) .. (0.77,-.77);
    \filldraw[orange] (-.77,-5.64) circle[radius=3pt] {};
    \filldraw[orange] (.77,-5.64) circle[radius=3pt] {};
    \draw[orange,thick, line width=2.5pt] (-0.77,-5.64) .. controls +(0.4, 0.5) and +(-0.4, 0.5) .. (0.77,-5.64);
    \draw[mydarkblue,thick, line width=2.5pt] (-2.4,-4) .. controls +(-0.1, -0.27) and +(0.4, -0.9) .. (-2.4,-2.45);
    \filldraw[mydarkblue] (-2.4,-2.45) circle[radius=3pt] {};
    \filldraw[mydarkblue] (-2.4,-4) circle[radius=3pt] {};
    \draw (0,-3.33) node[cross=5.5pt] {};
    \draw[dotted, line width=1pt] (0,-3.2) node[cross=5.5pt] {};
\end{tikzpicture}  
\caption{\small The $(r, \tau)$ section of a $4$-replica wormhole made of two copies of the $X_\mu$ manifold. An identical copy hides in the back, glued through the dust worldvolumes $W_i$, and a  ${\bf S}^{d-1}$ lies at each point. Similar geometries can be constructed replacing the $X_\mu$ manifold by the vacuum $X_0$ manifold.  } \label{fig:3}
\end{figure}
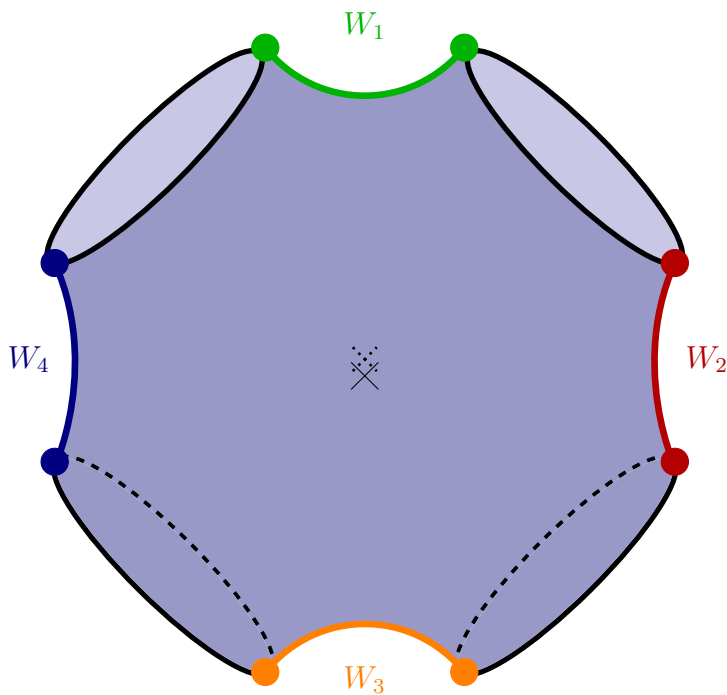

Classical saddle points meeting these boundary conditions can be constructed by picking two manifolds with thermal AdS asymptotic behavior, such as the Euclidean black hole manifolds $X_{\mu_L}$ and $X_{\mu_R}$, or their $\mu=0$ counterparts, each one missing $n$ portions cut out along the $n$ shell worldvolumes $W_i$ and glued back to back along the $W_i$ making up a sort of `sprocket' figure with  `front' and `back' faces corresponding to the left and right CFTs (\cref{fig:3}).  In principle, these front and back faces can have different geometries and even different topologies, since we can also construct wormholes by cutting and gluing vacuum $X_0$ manifolds. The sum rule for the range of $\tau$ coordinates is now 
\begin{equation}\label{sumrulew}
\sum_W \Delta \tau_{W} =\beta(\mu_L) - n\beta_L = \beta(\mu_R) - n \beta_R \;. 
\end{equation}
Incidentally, the $n$-wormhole manifolds can be generalized to arbitrary choices of the effective CFT inverse temperatures, different for each copy. In this case, the sum rule generalizes further to $\sum_i \Delta \tau_{W_i} =\beta(\mu_s) -\sum_j \beta_{s,j} $, where $s=L, R$ labels the front or back sides.  

To avoid terminological confusions, we should emphasize that there are two types of `wormhole' in each of the Euclidean manifolds $X_{i_1 \dots i_n}$, representing two different notions of connectivity. First, we have the Euclidean vestiges of the original Lorentzian wormholes, which connect the left and right `faces'  through the shell's worldvolumes $W_i$. In addition, we have the genuine `replica wormholes' which connect what microscopically would be a disconnected product of $n$ overlaps. This `replica-connectivity' is realized in the interior of each face, at radii $r< {\bar r}_i$, where ${\bar r}_i$ is the turning point of the $W_i$ submanifolds.

In the universality  limit $m\rightarrow \infty$ the Euclidean time extent of the dust  $\Delta \tau_{W_i}$ shrinks to zero as $\ell^2 /{\bar r}$  and the path integral (\ref{woh}) 
factorizes into a left-right pair, up to cutoff-dependent contact terms ${\rm C}_j$ at the endpoints of the $W_i$, which become local on the boundary and only depend on $m_j$, rather than the preparation inverse temperatures:
\begin{equation}\label{fact}
\overline {Z}_{i_1, \cdots i_n} \Big |_{m\to\infty} =\overline {Z} (n\beta_L) \;\overline {Z}(n\beta_R)\; \prod_{j=1}^n {\rm C}_j \;.
\end{equation}
 The corrections to this factorization rule scale as $(1/m)^{1 \over d-1}$ as $m\to \infty$. The function $\overline{Z}(\beta)$ is now an ordinary partition function over manifolds with canonical asymptotic
thermal AdS boundary conditions at inverse temperature $\beta$, i.e. the gravity approximation to the thermal partition function of a one-sided CFT.  Remarkably, we have
obtained precisely the GPI version of (\ref{prodi}), with each contact term ${\rm C}_j$ corresponding to the variance of the Gaussian distribution of the operator ${\cal O}_j$. \footnote{An averaging procedure which can reproduce the GPI expression at finite $m$ was proposed in \cite{Martinsolo}, where the operator matrix elements are chosen within  a generalized ETH {\it ansatz}, ${\cal O} \sim ({\rm Smooth}) \cdot ({\rm Random})$ and the smooth component is chosen to fit the classical shell dynamics.  See \cite{mdb} for more discussions of state averaging versus geometry. }

The authors of \cite{BLMS} show that the contact terms are universal when computed from the GPI and can be canceled by properly normalizing the canonical states. For the connected cyclic combination, this can be achieved by dividing (\ref{woh}) by the product of squared norms $\prod_{j=1}^n \overline{\langle \psi_{i_j} | \psi_{i_j }\rangle} $, that is to say
\begin{equation}\label{cic}
{\overline {\prod_{\rm cyclic} \langle \psi_{k_j} | \psi_{k_{j+1}} \rangle} \over \prod_j \overline{ \langle \psi_{k_j} | \psi_{k_j} \rangle}} \Bigg |_{m\to\infty}                 =               \left({\overline{Z} (n\beta) \over \overline{Z}(\beta)^n}\right)^2\;,
\end{equation}
where we have set $\beta_{L } = \beta_{R}= \beta$ in a further simplification of the involved formulae.

The l.h.s. of (\ref{cic})  determines the  `annealed' approximation to the GPI-averaged $n$-th moment of the Gram matrix.  Its most remarkable property is its universality, in the sense that it does not depend on the particular state indices entering the l.h.s. (this is precisely the reason why we refer to the $m\to\infty$ limit as the `universality limit'). Given the result (\ref{cic}) one can follow \cite{Stanford} and  derive a Schwinger--Dyson equation for the GPI approximation to the Gram resolvent ${\cal R}(\lambda) = (\lambda -{\cal G})^{-1}$, 
\begin{equation}\label{resol}
{\rm Tr} \,\overline{{\cal R}(\lambda)} = {{\cal N}_w \over \lambda} + \sum_{n=1}^\infty {1\over \lambda^{n+1}} \left({\overline{Z} (n\beta) \over \overline{Z}(\beta)^n}\right)^2 \,\left({\rm Tr} \,\overline {{\cal R}(\lambda)}\right)^n\;.
\end{equation}
This equation is easy to solve as a geometric series precisely when the r.h.s. of (\ref{cic}) is proportional to an $n$th power, say $a \,b^n$. In this case, the solution of (\ref{resol}) leads to  ${\rm rank} ({\cal G}) = a$. In particular, if we adopt a crude approximation of a fully degenerate  band of ${\cal N}$ states at energy $E$, we get 
$\overline{Z}(\beta)\sim {\cal N} \exp(-\beta E )$ and $\overline{Z}(n\beta) \sim {\cal N} \exp (-n\beta E )$, the quotient (\ref{cic})  resulting in $1/{\cal N}^{2(n-1)}$. With this input, the solution of (\ref{resol}) yields   
${\rm rank} ({\cal G}) = {\cal N}^2$, in agreement with the degenerate-band {\it ansatz} for the double-sided system.  

Taken at face value, this argument based on the extreme narrow-band approximation is  a mere consistency check. In fact, the form of the functions $\overline{Z}(n\beta)$ and $\overline{Z}(\beta)$, as they come from the GPI, is not quite compatible with the extreme narrow-band approximation. A more systematic treatment using inverse Laplace transforms to achieve the band projections was proposed in \cite{roberto}. In Section 3 of this note we will introduce a version of this formalism in some detail. 
We end this section with some observations regarding the stability of the classical saddle points entering the gravitational evaluation of (\ref{cic}), which motivate our discussion in the next section. 

\subsubsection*{Instabilities} 

Before addressing the microcanonical projection of (\ref{cic}), there is a rather general issue affecting the $n$-copy wormholes. Going beyond the classical approximation to the evaluation of GPI partition functions, one encounters fluctuation
determinants which may develop imaginary parts due to the occurrence of negative eigenvalues. This is quite clear precisely in the universality limit, since all wormhole partition functions factorize in this limit into left and right partition functions of standard thermal AdS manifolds. For these, the qualitative form of the partition function can be estimated on physical grounds. In terms of the effective action
\begin{equation}\label{effa}
I_{\rm eff} = -\log {\overline Z}[X] =  I[X] + {{1\over 2}} {\rm Tr} (-1)^F \log \,[\,I''\,]_{X}  + {\rm higher \; loops} \;.
\end{equation}
Here, $I[X]$ is the classical action of $X$, of order $1/G$. The  $I''$ occurring in the one-loop term, of order $G^0$,  are a set of fluctuation differential operators for all propagating perturbations on $X$ (bosonic and fermionic),   requiring  gauge-fixing and  special treatment for the gravitational conformal mode, as well as all zero (or quasi-zero) modes and possible negative eigenvalues of $I''$.

There are non-extensive terms, with logarithmic dependence on $\ell$,  in the one-loop effective action coming from quasi-zero modes. Translational zero modes are lifted by the AdS background curvature and the associated position-dependence of the local temperature. However, in the limit that $\beta \ll \ell$ we can have a small black hole with approximate translational modes in an effective box of size $\ell$. This contributes  to the partition function a factor $\ell^d\, (M(\beta) /2\pi \beta)^{d/2}$, just like a non-relativistic particle of mass $M(\beta)$ in $d$-dimensional box of size $\ell$. Equivalently, by raising this factor to the exponent (\ref{effa}), we get one of the sources of logarithmic corrections in the effective action. Another one comes from the renormalization of logarithmic UV divergences, contributing to conformal anomalies in even dimensions, such as the Euler density counterterm in four dimensions (see \cite{sen} for a general discussion of such logarithmic terms). 

One piece of the one-loop effective action which is easy to estimate is the extensive contribution of massless fields. We can approximate this by summing the free energies of massless fields at local thermal equilibrium with temperature $1/\beta_{\rm loc}$, where $\beta_{\rm loc} = \beta \sqrt{g_{\tau\tau}}$. This results in an expression of the form
\begin{equation}\label{optical}
I_{\rm rad} (X) \approx -n_*\, a_{d+1} \, \int_X d{\rm Vol}_{\rm reg} \,(\beta_{\rm loc})^{-d-1} = -n_* \,a_{d+1} \, \beta^{-d-1} \,\widetilde{ \rm Vol}_{\rm reg}(X)\;,
\end{equation}
where $a_D= \Gamma( D/2) \,\zeta(D)\, \pi^{-D/2}$ is the  radiation constant in $D$ spacetime dimensions and $n_*$ is the effective number of thermal species.  In this estimate,  ${\widetilde {\rm Vol}}_{\rm reg}$ is the regularized optical volume,\footnote{Optical volumes are useful because they express naive extensivity for radiation in equilibrium (see \cite{optical} and \cite{opticalm}).} measured with respect to the conformally rescaled metric ${\widetilde g}_{\mu\nu} = g_{\mu\nu} / g_{\tau\tau} = g_{\mu\nu} /f(r)$. It is regarded as `regularized' because one must subtract the
near-horizon divergence, from the vanishing of $g_{\tau\tau}$ when there is a horizon. This divergence is counted as part of the classical term $I(X)$, through the renormalization of Newton's constant  \cite{su}. The crucial aspect of (\ref{optical}) for our purposes is the finite optical volume of asymptotically AdS regions, so that $I_{\rm rad} \sim -(\ell /\beta)^d$, as long as $\beta \ll \ell$. 

Beyond the free radiation approximation, the graviton contribution to $I_{\rm rad}$ develops a dynamical instability at high temperatures, coming from a one-loop generated tachyonic mass for the time-time component of the graviton, as explained in the classic paper  \cite{gpy}. In the regime $\beta \ll \ell$,  it has the form 
$m_g^2 \sim - G /\beta^{d+1}$.  This is the relativistic version of the classic Jeans instability and it occurs when the Jeans length $\ell_J =\sqrt{ \beta^{d+1} /G}$ becomes smaller than the `box size' $\ell$. Physically, it corresponds to the radiation at temperature $1/\beta$ collapsing into black holes of size $\ell_J$.  At the level of the partition function, all momentum modes with energy between
$\ell^{-1}$ and $\ell_J^{-1}$ become tachyonic, each one contributing a factor of $1/i$ to the partition function. 

Finally, for manifolds of $X_{\mu^-}$ type, having a small black hole with $\beta(\mu) \ll \ell$, the spin-two, transverse-traceless sector of $I''$ develops a negative eigenvalue \cite{gpy}. This mode is well-localized near the horizon and the negative eigenvalue scales as $\lambda_{\rm GPY} \sim -1/\beta^2$. In the case of AdS black holes, $\lambda_{\rm GPY}$ stops being negative precisely at the maximum of the function $\beta(\mu)$, so that the GPY instability coincides with the region of negative specific heat \cite{prestidge}.  The contribution to the partition function gives a factor of $1/\sqrt{\lambda_{\rm GPY}} \sim 1/i$, where we are taking the positive branch of the square root, consistent with the fact that the GPY negative eigenvalue migrates from an ordinary positive eigenvalue as 
the mass parameter is continued from the region of large AdS black holes into the region of small AdS black holes. 

\begin{figure}[htpb!]
\centering
\begin{tikzpicture}[scale=1.15]
\def\a{.7} 
\def\b{.25} 
    \filldraw[myblue, decorate, decoration={random steps,segment length=3pt,amplitude=1pt}]  (5,0) circle ellipse({0.75} and {4});
    \filldraw[white]  (4.3,-1).. controls +(0.3, 0.3) and +(0.3,-0.3) .. (4.3,1)-- (4.1,1)-- (4.1,-1)--(4.3,-1)--cycle;
    \draw[mydarkblue,thick, line width=2.5pt] (4.3,-1).. controls +(0.2, 0.3) and +(0.2,-0.3) .. (4.3,1); 
    \filldraw[mydarkblue] (4.3,1) circle[radius=1pt] {};
    \filldraw[mydarkblue] (4.3,-1) circle[radius=1pt] {};
    \draw{} (3.9,1.2)circle (0pt) node[mydarkblue]{$W_4$};
    \filldraw[white,line width=0.1](4,.7).. controls +(0.5, 0.1) and +(-0.1,-1) ..(4.70,2)--(4.70,-2).. controls +(-0.1,1) and +(0.5, -0.1) ..(4,-.7)--cycle;
    \filldraw[myblue] (4,.7).. controls +(-1, 0) and +(0.5, 0.2) ..(-3.5,0.5).. controls +(-0.5, -0.2) and +(-0.5, 0.2) ..(-3.5,-0.5).. controls +(0.5, -0.2) and +(-1, 0).. (4,-.7).. controls +(0.5, -0.1) and +(-0.1, 1) ..(4.7,-2)--(4.7,2).. controls +(-0.1, -1) and +(0.5,0.1).. (4,.7)--cycle;
    \draw[line width=0.8] (4,.7).. controls +(-1, 0) and +(0.5, 0.2) ..(-3.5,0.5);
    \draw[line width=0.8] (4,-.7).. controls +(-1, 0) and +(0.5, -0.2) ..(-3.5,-0.5);
    \draw[line width=0.8] (-3.5,0.5).. controls +(-0.5, -0.2) and +(-0.5, 0.2) ..(-3.5,-0.5);
    \draw[line width=0.8] (4,.7 ).. controls +(0.5, 0.1) and +(-0.1,-.8) ..(4.7,2);
    \draw[line width=0.8] (4,-.7).. controls +(0.5, -0.1) and +(-0.1, 1) ..(4.7,-2);
    \draw[dashed] (4,0) ellipse({\b} and {\a});
    \draw[] (3.8,0) circle (0pt) node[anchor= south east] {$\beta$};
    \draw[<->] (4.2,-1.5) --(-3.5,-1.5) ;
    \draw[] (0.5,-1.5) circle (0pt) node[anchor= north ] {$\ell$};
    \draw[] (-4.1,0) circle (0pt) node[anchor= east] {$r_h$};
    \filldraw[white](5.75,-1).. controls +(-0.3, 0.3) and +(-0.3, -0.3) .. (5.75,1)--(5.9,1)--(5.9,-1)--(5.75,-1)--cycle;
    \draw[myred,thick, line width=2.5pt] (5.75,-1).. controls +(-0.3, 0.3) and +(-0.3,-0.3) .. (5.75,1); 
    \filldraw[myred] (5.75,1) circle[radius=1pt] {};
    \filldraw[myred] (5.75,-1) circle[radius=1pt] {};
    \draw{} (6,0)circle (0pt) node[myred]{$W_2$};
    \filldraw[white,thick, line width=2.5pt] (4.6,-3.3) .. controls +(0.2, 0.2) and +(-0.2, 0.2) .. (5.4,-3.3)--(5.4,-4)--(4.6,-4)--(4.6,-3.3)--cycle;
    \filldraw[orange] (5.4,-3.3) circle[radius=1pt] {};
    \filldraw[orange] (4.6,-3.3) circle[radius=1pt] {};
    \draw[orange,thick, line width=2.5pt] (4.6,-3.3) .. controls +(0.2, 0.2) and +(-0.2, 0.2) .. (5.4,-3.3);
    \draw{} (5,-3.7)circle (0pt) node[orange]{$W_3$};
    \filldraw[white,thick, line width=2.5pt] (4.6,3.3) .. controls +(0.2, -0.2) and +(-0.2, -0.2) .. (5.4,3.4)--(5.4,4)--(4.6,4)--(4.6,3.3)--cycle;
    \filldraw[mygreen] (5.4,3.3) circle[radius=1pt] {};
    \filldraw[mygreen] (4.6,3.3) circle[radius=1pt] {};
    \draw[mygreen,thick, line width=2.5pt] (4.6,3.3) .. controls +(0.2,- 0.2) and +(-0.2, -0.2) .. (5.4,3.3);
    \draw{} (5,3.7)circle (0pt) node[mygreen]{$W_1$};
\end{tikzpicture}
\caption{An Euclidean `flat box'  corresponding to a small Euclidean black hole which supports a GPY mode, localized near the `tip of the cigar'.}
\label{fig:4}
\end{figure}
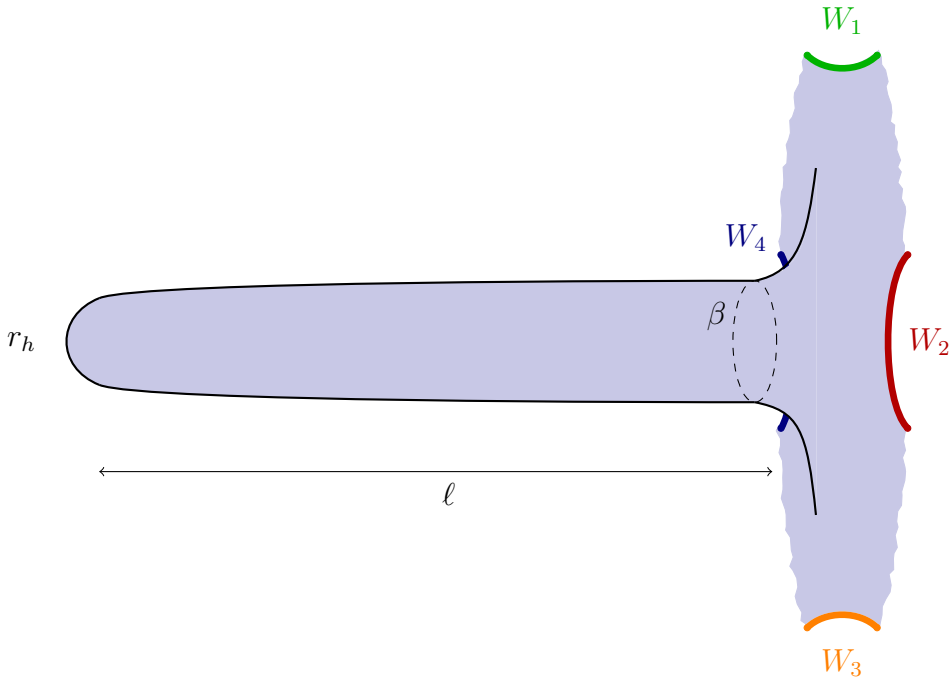

Both the Jeans and the GPY instability are characteristic of Euclidean manifolds with a large `flat space box', corresponding to high temperatures $\beta \ll \ell$ (see \cref{fig:4}) and $d>2$. The Jeans instability afflicts both the small black-hole manifold $X_{\mu^-}$ and the vacuum thermal one $X_0$. It can be controlled by maintaining the effective box size well within the Jeans length, $\ell < \ell_J$. In contrast, the GPY instability affects the small black hole manifold $X_{\mu^-}$ specifically and it cannot be removed by tuning parameters.  
At any rate, if these instabilities 
exist for the strict universality limit, they will persist away from the limit, so long as the dust shells remain far from the flat box, i.e. ${\bar r} \gg \ell$. 

What are we to make of these instabilities regarding the estimate of state overlaps? In \cite{gpy}, the imaginary factors contributing to $\overline {Z} [X_{\mu^-}]$ are regarded as computing the imaginary part of the thermal free energy of {\it flat} space, after a dilute-gas resummation, i.e. a decay width. In particular, the small black hole manifold $X_{\mu^-}$ is interpreted as a kind of gravitational sphaleron configuration, mediating the decay of the thermal state in flat space, through thermal excitation over a barrier in the space of metrics.  On the other hand, the use of Euclidean GPIs to estimate norms and inner products of states does not easily tolerate 
the occurrence of complex phases. For instance, the formally defined partition function  $\overline{Z}[X_{\mu^-}]$ on a `small black hole' manifold is proportional to $1/i$, so that $\overline{Z}[X_{\mu^-, m, \mu^-}]$ in (\ref{epi})   is proportional to $(1/i)^2 = -1$ for large $m$, and it cannot be 
interpreted as a bulk approximation to a state norm-squared,  despite the fact that the exact CFT quantity (\ref{inpet}) is positive-definite by construction. The resolution is of course the Hawking--Page transition, namely the manifold $X_{\mu^-}$ is never the dominant saddle point for any value of the preparation inverse temperatures. For any solution
of the sum rule (\ref{sumrp}) constructed by gluing $X_{\mu^-}$ manifolds, the stable `large black hole' manifold constructed from $X_{\mu^+}$ also satisfies the constraints and has a larger value of the partition function. 

For another, less elating consequence of the Hawking-Page transition, consider the central result (\ref{cic}). Let us suppose that $\beta$ is small enough so that  $\overline{Z}(\beta)$ in the denominator is dominated by a large AdS black hole $X_{\mu^+}$. No matter how small $\beta$ may be, eventually, we will have $n\beta \gg \ell$ for sufficiently large $n$. Therefore,
for all but a finite number, the quotients in (\ref{cic}) are dominated by the $X_0$ manifold at inverse temperature $n\beta$ in the numerator, and the $X_{\mu^+}$ manifold in the denominator. This lack of coherence of the saddle points prevents a simple interpretation of the canonical PETS overlaps as related to the entropy of a particular black hole.

In the rest of this note we show how all these issues can be resolved within the microcanonical framework, including the potential problems posed by the  one-loop instabilities afflicting 
manifolds with large flat-space portions.

\section{Microcanonical Overlaps and their Stability}

\noindent

\subsubsection*{The Microcanonical States}
\noindent

A microcanonical version of the PETS states is simply obtained by applying an inverse Laplace transform from inverse temperature to energy variables
\begin{equation}\label{MPETO}
|\psi_{E_L , {\cal O}, E_R} \rangle =  \int_{\gamma_L} {1\over 2} \,{d\beta_L \over 2\pi i} \int_{\gamma_R} {1\over 2}\,{d\beta_R \over 2\pi i} \; e^{\beta_L E_L/2 + \beta_R E_R /2}\;\; |\psi_{\beta_L, {\cal O}, \beta_R} \rangle\;,
\end{equation}
where $\gamma_{L,R}$ are contours running  upwards, parallel to the imaginary axis, in the complex $\beta_{L,R}$ planes, and the various factors of $1/2$ are introduced for notational convenience in what follows.  Inserting the explicit form of the PETS states we find
\begin{equation}\label{MPETOo}
|\psi_{E_L , {\cal O}, E_R} \rangle =\sum_{n,m} | n^* \rangle_L \,\delta(E_L - E_{L,n}) \,{\cal O}_{n\,m} \,\delta(E_R - E_{R,m}) \,|m\rangle_R\;.
\end{equation}
These states have distributional coefficients. It is convenient to smear them out with a filter window function, producing the states
\begin{equation}\label{filter}
|{\tilde \psi}_{\cal O} \rangle = \int dE_L F(E_L) dE_R F(E_R) \,| \psi_{E_L, {\cal O}, E_R} \rangle \;,
\end{equation}
whose components in the orthonormal basis $|n^* \rangle_L \otimes |m \rangle_R $ are 
\begin{equation}\label{gore}
{\tilde \psi}_{nm} = F(E_n^L) \,{\cal O}_{n\,m} \,F(E_m^R)\;.
\end{equation}

The general inner product is given by 
\begin{equation}\label{inn}
\langle {\tilde\psi}_{\cal O} | {\tilde \psi}_{\cal O'}\rangle =  \sum_{m,n} F(E_{L,n})^2 \,{\cal O'}_{n\,m} \,F(E_{R,m})^2 \,{\cal O}^\dagger_{\,m\,n} \;.
\end{equation}
We can now undo the previous steps to obtain an expression involving a canonical one-sided trace and  only two (instead of four) contour integrals:
\begin{equation}\label{undo}
\langle {\tilde\psi}_{\cal O} | {\tilde \psi}_{\cal O'}\rangle =\left( \prod_{s=L,R}  \int dE'_s F_s (E'_s)^2 \int_{\gamma_s}{d\beta_s \over 2\pi i} \;e^{\beta_s \,E'_s}\right) \,{\rm Tr}\left({\cal O}^\dagger e^{-\beta_L H} {\cal O'} e^{-\beta_R H} \right)\;.
\end{equation}
The canonical one-sided trace is now well suited for reading off path-integral expressions in bulk fields. 

In principle, we have considerable freedom in specifying the left and right window functions $F_s (E)$. For instance, one choice that works well with the calculation of Laplace transforms by saddle-point approximations is the use of Gaussian profiles. On the other hand, for microscopic computations, it is crucial to emphasize the finite dimensionality of the band. Thus a sharp pass function, supported on the  left and right bands  $B_s$,  will be more appropriate, 
\begin{equation}\label{passf}
F_s (E) = \Theta (E- E_s + \Gamma/2) - \Theta(E-E_s - \Gamma/2)\;, 
\end{equation}
where $\Theta$ is the standard Heaviside function. In particular $F_s (E)^2 = F_s (E)$ for this choice. This is equivalent to projecting directly the operators onto the band, $\pi_L {\cal O} \pi_R $, namely  we have 
\begin{equation}\label{projo}
\langle {\tilde\psi}_{\cal O} | {\tilde \psi}_{\cal O'}\rangle  = {\rm Tr} \left(\pi_R{\cal O}^\dagger \, \pi_L \,{ \cal O}' \,\pi_R\right)\;,
\end{equation}
where $\pi_s$ is the orthogonal projector onto the band $B_s$. 

We can now adapt the microscopic coarse-graining (\ref{microcg}) to the projected operators
\begin{equation}\label{microcgb}
\overline{ (\pi_L {\cal O}_j \pi_R)_{l\,r}  (\pi_R {\cal O}_k^\dagger \pi_L)_{r' \,l'} }  \Bigg |_{{\cal O}-{\rm average}}= {\rm C}_j \,\delta_{j\,k}\;\delta_{r\,r'} \,\delta_{l\,l'}\;,
\end{equation}
where the indices $r, r'$ and $l, l'$ are now restricted to the right (respectively left) band. Inserting this averaging rule into the microcanonical version of the cyclic product we find
\begin{equation}\label{cyc}
\overline{\prod_{\rm cyclic} \langle {\tilde \psi}_{i_j} | {\tilde \psi}_{i_{j+1}}\rangle} \,\Big |_{{\cal O}-{\rm average}} =\left( \prod_{j=1}^{{\cal N}_w} {\rm C}_j \right)\; {\cal N}_\Gamma (E_L) \,{\cal N}_\Gamma (E_R)\;,
\end{equation}
where ${\cal N}_\Gamma (E_s)$ are the dimensions of the $B_s$ bands. 
The result for the annealed estimate of the cyclic product of Gram matrix elements is 
\begin{equation}\label{gramsci}
{\overline{\prod_{\rm cyclic} \langle {\tilde \psi}_{i_j} | {\tilde \psi}_{i_{j+1}}\rangle} \over \prod_j \overline{\langle {\tilde \psi}_{i_j} | {\tilde\psi}_{i_j} \rangle}} \;\Bigg |_{ {\cal O}-{\rm average}} = \left({1\over {\cal N}_\Gamma (E_L) \, {\cal N}_\Gamma (E_R)}\right)^{n-1}\;,
\end{equation}
which ensures the correct expected  result for the Gram's matrix rank, i.e. ${\rm rank}({\cal G}) =  {\cal N}_\Gamma (E_L) \, {\cal N}_\Gamma (E_R)$.

These results are not surprising when one notices that the components in (\ref{gore}) furnish a  rectangular matrix of ${\cal N}_w$ rows and ${\cal N}_{\rm B}$ columns  
$
{\cal A}^j_{J} = {\tilde \psi}^j_{l\,r} = ({\cal O}_j)_{l\,r}
$, where we interpret the pair $(l\,r)$ as a single double index, $J$,  running over the combined dimension of the band ${\cal N}_{\rm B} =  {\cal N}_\Gamma (E_L) \, {\cal N}_\Gamma (E_R)$. 
Then, ${\cal A} {\cal A}^\dagger$ is an ${\cal N}_w \times {\cal N}_w$ matrix of inner products which is equivalent to the Gram matrix, up to normalization of the states. In particular, if the individual ${\cal A}$-columns are drawn as independent, normally distributed vectors of variance ${\rm C}_j$, the ${\cal A}$ matrix will have maximal rank, equal to ${\cal N}_{\rm B}$ with probability approaching one in the limit of large matrices. Therefore, ${\cal A} {\cal A}^\dagger$ will also have rank ${\cal N}_{\rm B}$, no matter how large ${\cal N}_w$ may be. More technically,  the matrix ${\cal A} {\cal A}^\dagger$ is distributed according to the so-called Wishart ensemble \cite{Wishart}. In the limit of large matrices, it is a standard result that the Wishart distribution leads to an averaged resolvent whose Schwinger--Dyson equation of type (\ref{resol}) is summable as a geometric series. The associated density of eigenvalues has the so-called Marchenko--Pastur form, and has the property  ${\rm rank} ({\cal A} {\cal A}^\dagger) = {\cal N}_{\rm B}$ as ${\cal N}_w > {\cal N}_{\rm B}\gg 1$, (see \cite{muck} for a recent discussion of this distribution in the black-hole context). 

\subsubsection*{The Microcanonical GPI Overlaps}
\noindent

We now check that the same results follow from the GPI-induced coarse-graining. We start from the GPI version of the cyclic product 
 \begin{equation}\label{cyme}
\overline {\prod_{\rm cyclic} \langle {\tilde \psi}_{i_j} | {\tilde \psi}_{i_{j+1}} \rangle} 
 =  \left( \prod_{s=L,R}  \int_{\vec B_s}  d{\vec E}_{s} \right)\; \overline{ \Omega}_{i_1 \dots i_n} ({\vec E}_{s}) \;,
\end{equation}
where we have adopted a condensed  vector notation for the $n$-replicated variables. In particular, we denote ${\vec B}_s$ the left and right hypercubes in the $n$-replicated energy space. These hypercubes are centered at the vector $(E_s, E_s, \dots, E_s)$ and have sides of length $\Gamma$. $\overline{\Omega}_{i_1 \dots i_n}$ is a sort of $n$-variable   generalization of the standard $n=1$  density of states:
\begin{equation}\label{gendens}
\overline{\Omega}_{i_1 \dots i_n} ({\vec E}_{s}) = \left( \prod_{s= L, R}  \int_{{\vec \gamma}_{s}} {d{\vec \beta}_{s} \over (2\pi i)^n} \,e^{{\vec \beta}_{s} {\vec E}_{s}}\right) \,{\overline Z}_{i_1 \dots i_n} ({\vec \beta}_s, {\vec m})  \;,
\end{equation}
obtained by multiple Laplace transforms of the same generalized partition function appearing in (\ref{woh}), but continued to arbitrary values of the preparation temperatures $\beta_{j,s}$ 
in each ${\bf S}^1$ factor of the conformal boundary. Applying now the universality limit (\ref{fact}) we obtain, up to the contact terms, a left-right product of the non-trivial integral
\begin{equation}\label{keyint}
\overline{\Omega}_n  =\int_{{\vec \gamma}} {d{\vec \beta} \over (2\pi i)^n}  \,e^{{\vec \beta} {\vec E}} \,\overline {Z}({\vec 1} \cdot {\vec \beta})\;.
\end{equation}
where ${\vec 1} = (1, 1, \dots, 1)^t$ is the column $n$-vector of unit entries, so that ${\vec 1} \cdot {\vec \beta} = \sum_j \beta_j$. 
To evaluate this integral, we consider the change of variables  ${\vec \beta} \to {\vec \beta}' = J {\vec \beta}$, where $J$ is any orthogonal matrix that takes the column unit vector $(1/\sqrt{n}, 1/\sqrt{n}, \dots, 1/\sqrt{n})^t $ into the column unit vector $(1, 0, \cdots, 0)^t$. The first row of this matrix is determined to be $J_{1j} = 1/\sqrt{n}$, and acting on a vector of components $v_j$ it generates a vector of components $v'_j$ such that $v'_1 = (\sum v_j) /\sqrt{n}$. Since the matrix $J$ is orthogonal, it leaves invariant the integration measure $\prod_j dv_j$, as well as the scalar products.  Therefore we find 
 \begin{equation}\label{omega}
\overline{\Omega}_n = \left( \prod_{j>1} \delta(E'_j) \right)\int_{\gamma'_1} {d\beta'_1 \over 2\pi i} \,e^{\beta'_1 E'_1} \,\overline{Z}(\sqrt{n} \beta'_1) = \left(\prod_{j>1} \delta (E'_j)\right) \;{1 \over \sqrt{n}} \,\overline{\Omega} (E'_1 /\sqrt{n})\;, 
\end{equation}
where $\overline{\Omega}(E)$ is the standard one-sided density of states as computed from GPI approximation to the canonical partition function. It remains now to integrate
this expression over $J(B_s)$,  the hypercubes of energy bands, rotated by the $J$ transformation. The delta functions of the $E'_{j>1}$ variables localize the integral over the one-dimensional  intersection of the rotated $E'_1$ axis with the band hypercube.  In the rotated frame, the $E'_1$ axis cuts the hypercube along its main diagonal, which has length $\Gamma \,\sqrt{n}$. Therefore, a change of variables $E'_1 \to E'_1 /\sqrt{n}$ restores this interval to length $\Gamma$. We thus conclude that the final answer, after dividing by the normalizations, is the
expected one:
\begin{equation}\label{prelim} 
{\overline{\prod_{\rm cyclic} \langle {\tilde \psi}_{i_j} | {\tilde \psi}_{i_{j+1}}\rangle}\over \prod_j \overline{\langle {\tilde \psi}_{i_j} | {\tilde\psi}_{i_j} \rangle} }\;{\Bigg |}_{m\to \infty} = \left({1\over {\bar{\cal N}}_\Gamma (E_L) {\bar {\cal N}}_\Gamma (E_R)}\right)^{n-1}\;,
\end{equation}
where 
\begin{equation}\label{nbar}
{\bar {\cal N}}_\Gamma (E) \equiv  e^{S_\Gamma (E)} \equiv \int_{E - {\Gamma \over 2}}^{E + {\Gamma \over 2}} dE' \; \overline{ \Omega} (E')
\end{equation}
is the GPI estimate of the band dimension. It is interesting that this result depends in a somewhat detailed way on the treatment of the energy bands. Had we used Gaussian window functions instead of sharp projectors in (\ref{passf}), the numerator of the l.h.s. of  (\ref{prelim}) would present a narrower effective band thickness, offset by a factor  of $\sqrt{n}$ with respect to the denominator. Thus the result of the r.h.s. would have required an {\it ad hoc} rescaling of the windows when computing the numerator.

\subsubsection*{Semiclassical Stability }
\noindent
The result (\ref{prelim}) gives the expected answer for any choice of microcanonical window. In particular, we may consider zooming into the region of the spectrum where
the potentially unstable manifolds $X_0$ and $X_{\mu^-}$ are supposed to make a dominant contribution. In this situation, it is interesting to ask whether the complex phases associated to these instabilities may ruin the formal evaluation leading to (\ref{prelim}). The threatening instabilities arise in high bulk-temperature situations, $\beta \ll \ell$, precisely in the universality limit, for which we may neglect the effect of the dust shells. We are thus led to the standard analysis of GPI calculations of the density of states, as described for instance in the classic paper \cite{HP}. 

Starting from the expression (\ref{gendens}) we can write ${\vec E}_s = {\vec \omega}_s + {\vec 1}\,E_s$, where ${\vec \omega}_s$ are variables confined to the hyper-cubic bands translated to the origin, ${\vec B}_s(0)$.  The microcanonical cyclic product can be expressed as 
\begin{equation}\label{smearL}
 \overline {\prod_{\rm cyclic} \langle {\tilde \psi}_{i_j} | {\tilde \psi}_{i_{j+1}} \rangle}  = \prod_{s=L, R} \int_{{\vec B}_s (0)} d{\vec \omega}_s\int_{{\vec \gamma}_s} {d{\vec \beta}_s \over (2\pi i)^n}  \,e^{{\vec \beta}_s\cdot {\vec \omega}_s}\;e^{\beta_s E_s}  \;\overline {Z}({\vec \beta}, {\vec m})\;,
\end{equation}
where we denote $\beta_s = {\vec \beta}_s \cdot {\vec 1} = \sum_j \beta_{s, j}$, the crucial integration variables, coupling to the central band energies $E_s$, and also controlling the partition function in the universality limit. This form of the integral suggests applying the same change of variables as before,  ${\vec \beta} \to {\vec \beta}' = J({\vec \beta})$, and ${\vec \omega} \to {\vec \omega}' = J({\vec \omega})$,  with $\beta_s = \beta'_{s,1} \sqrt{n}$, to obtain 
\begin{equation}\label{sprime} 
 \overline {\prod_{\rm cyclic} \langle {\tilde \psi}_{i_j} | {\tilde \psi}_{i_{j+1}} \rangle}  = \prod_{s=L, R} \int_{J({\vec B}_s (0))} d{\vec \omega'}_s\int_{{\vec \gamma}'_s} {d{\vec \beta}'_s \over (2\pi i)^n}  \,e^{{\vec \beta}'_s\cdot {\vec \omega}'_s}\;e^{\beta_s E_s}  \;\overline {Z}({\vec \beta}', {\vec m})\;.
\end{equation}
As indicated above, we are mostly interested in the dependence of the partition function on $\beta_s = \sqrt{n} \beta'_{s,1}$, which is in any case the only relevant variable in the universality limit. Given this mild dependence of the integrand on the `transverse' variables $\beta'_{s, j>1}$, it is then reasonable to simplify (\ref{sprime}), replacing the partition function by an average value in the transverse directions, say its value at the `replica-symmetric' point $\overline{Z}_{\rm sym} = \overline {Z}|_{\beta'_{j>1} =0}$, times a multiplicative  $O(1)$ factor. In doing so,  the integral over all the transverse inverse temperatures and energies  evaluates to one, and the remaining integral over the `longitudinal' variables is 
\begin{equation}\label{remas}
\prod_{s= L,R} \int_{-\Gamma \sqrt{n} / 2}^{\Gamma \sqrt{n} / 2} d\omega'_{s,1} \int_{\gamma'_s}{d\beta'_{s,1} \over 2\pi i} \;e^{\beta'_{s,1} \omega'_{s,1}} \; e^{\beta_s E_s}\; \overline{Z}(\beta_s, {\vec m})_{\rm sym}\;.
\end{equation}
Just as before, the $\sqrt{n}$ factors in the limits of integration result from the fact that $\omega'_1$ runs along the diagonal of the 
centered band hypercube. Changing variables from $\beta'_{s,1}$ to $\beta_s$ and doing the final energy integral we find, up to $O(1)$ numerical factors
\begin{equation}\label{simpleone}
 \overline {\prod_{\rm cyclic} \langle {\tilde \psi}_{i_j} | {\tilde \psi}_{i_{j+1}} \rangle}  \sim \prod_{s=L, R} \int_{ \gamma_s} {d \beta_s \over 2\pi i}  \,f_\Gamma (\beta_s)\;e^{\beta_s E_s}  \;\overline {Z}( \beta_s, {\vec m})_{\rm sym}\;,
 \end{equation}
 with the function 
$$
f_{\Gamma} (\beta) = {2\over \beta} \sinh (\beta\, \Gamma /2) 
$$
  introducing a smearing of the Laplace transform.  In evaluating the $\beta_s$ integral by saddle-point approximation for large $E_s$, we anticipate that the integrand will present a much stronger variation in the vicinity of the saddle point than $f_\Gamma$ itself. Therefore, we can just approximate the smearing function by its value at the saddle point. As long as the  band width $\Gamma$ is chosen much smaller than the physical temperatures of interest in the system,  we will always have ${\rm Re}(\beta_s) \cdot \Gamma \ll 1$ around the saddle point, and we may simply replace   $f_\Gamma |_{\rm saddle}$ by $\Gamma$.

We shall study (\ref{simpleone}) in a double saddle-point expansion. First, the canonical partition function  is defined as a sum of  perturbative partition functions around  GPI saddles, i.e.  the $n$-wormhole manifolds $X_{i_1, \dots, i_n}$. For each such saddle manifold, this defines a real function of $\beta_s$ which is analytically continued into the complex plane. This analytic continuation is subsequently integrated over $ \beta_s$, again by saddle-point approximation. In this last step,
we must choose a steepest-descent contour through the dominant saddle point. 

For a given saddle manifold $X$ of the GPI, candidate saddle points of the $\beta_s$ integral satisfy 
\begin{equation}\label{sad}
E_s + \partial_{\beta_s} \log \overline{Z}_X (\beta_s) \Big |_{\beta_s= \beta(E_s)} =0\;.
\end{equation}
The inverse-temperature sum rule for an $n$-replica wormhole manifold reads $\beta(\mu_s) = \beta_s + \sum_W \Delta \tau_{W}$, where $\beta(\mu_s)$ is the bulk $\tau$-identification period. Therefore, by shifting the integration variable in (\ref{simpleone}) by $\beta_s \rightarrow \beta_s - \sum_W \Delta \tau_W $, the saddle points in the new variable will coincide with
the bulk inverse temperatures of the saddle manifold, at the cost of introducing a prefactor $\exp(-E_s \,\sum \Delta \tau_W)$ in (\ref{simpleone}) which can be neglected in the universality limit.

With one-loop accuracy around each saddle manifold $X$,  the dominant saddle point for each value of $E_s$ will maximize the quantity
$
\exp\left(\beta_s E_s - I(X) - I_{\rm rad} (X)\right) 
$, 
where $I(X)$ is the classical Euclidean action of the saddle manifold $X$ and $I_{\rm rad} (X)$ is the radiation free energy of $X$, according to (\ref{optical}). Equivalently,
one must select the manifold which maximizes the entropy. 

\begin{figure}[htbp]
   \centering
   \includegraphics[width=4in]{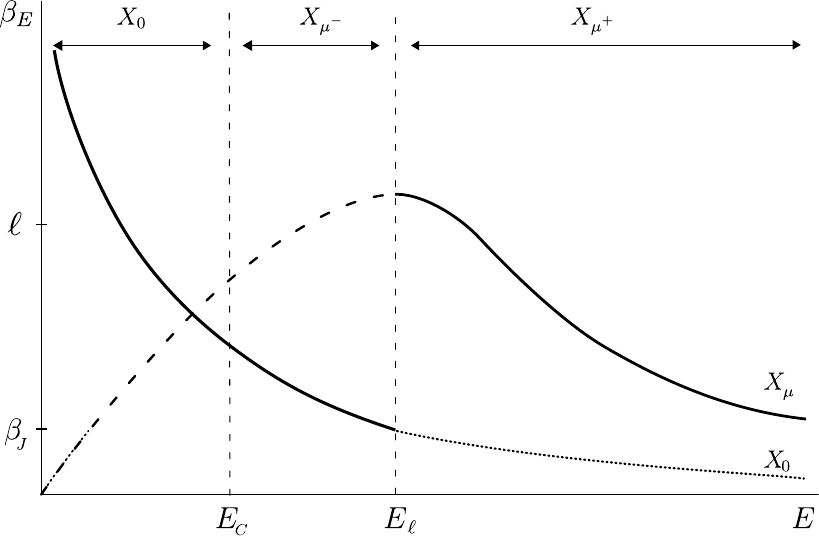}
   \caption{\small The bulk inverse temperature $\beta_E$  for thermal states on the manifolds $X_0$ and $X_\mu$, as a function of the microcanonical energy $E$. Regions of local instability are indicated by  discontinuous graph curves. The Jeans instability (dotted line) occurs for $\beta < \beta_J$ on both the $X_{\mu^-}$ and the $X_0$ curve. The GPY instability (dashed line) occurs for the monotonically increasing portion of the $X_\mu$ curve.  We also show the regions of dominance of the different manifolds.} \label{fig:5}
   \end{figure}

Taking as a guide the behavior in the strict universality limit, which reduces the problem to that of two decoupled (left and right) thermal AdS problems, we can expect that the $\beta(E_s)$ functions for the competing saddle manifolds will have the qualitative form shown in \cref{fig:5}.\footnote{It should be noted that this is a crude description based on our reduced cast of characters regarding degrees of freedom. More complete descriptions include localized black holes in higher dimensional compact factors and Hagedorn phases of long strings (see for example \cite{Thresholds}).} The inverse temperature of the $X_0$ manifold is completely determined by the radiation and scales like 
$
\beta_{X_0} \sim \ell /(\ell E)^{1/( d+1)}
$, 
while the inverse temperature of the black hole manifold $X_\mu$ is dominated by the $O(1/G)$ classical action and has different behavior according to the `large' or `small' black hole branches. These two branches merge at energies of order $E_\ell \sim N_* /\ell$, where $N_* \sim \ell^{d-1} /G$ is of the order of the central charge of the dual CFT.  For large black holes we have $\beta_{X^+} \sim \ell (N_* / \ell E)^{1/d}$, whereas the `small' black hole branch yields $
\beta_{X^-} \sim \ell (\ell E /N_*)^{1 /( d-2)} $. Both the small black hole manifold $X_{\mu^-}$ and the vacuum $X_0$ manifolds are Jeans-unstable for $\beta < \beta_J \sim \ell / N_*^{1 /( d+1)}$, and  the $X_{\mu^-}$ manifold is afflicted by the GPY instability along all its parameter space.

The two $\beta_{X_0}$ and $\beta_{X_\mu}$ curves cross around (actually, slightly below by an $O(1)$ factor), the `coexistence' energy, $E_C \sim N_*^\nu /\ell$, with $\nu = (d+1)/(2d-1)$. At this energy, the entropy of the $X_{\mu^-}$ manifold begins dominating over that of the $X_0$ manifold. Subsequently, around the threshold $E_\ell$, the $X_{\mu^-}$ manifold
continuously morphs into the $X_{\mu^+}$ manifold. We conclude that  the Jeans instability does not affect any of the dominant manifolds, but the GPY instability afflicts the dominant
manifold $X_{\mu^-}$ in the window $E_C < E < E_\ell$.

Summarizing, in the strict universality limit, the factorized $n$-replica wormhole will feature an $X_0$ manifold at its core for $E_s<E_C$, with no local instabilities and positive heat capacity. In the interval $E_C < E_s < E_\ell$ we have $X_{\mu^-}$ manifolds  at the core with negative heat capacity and one negative GPY eigenvalue. Finally, for $E_s > E_\ell$ we have $X_{\mu^+}$ manifolds at the core  with both local and thermodynamic stability. 

This qualitative picture is expected to hold away from the strict universality limit, so long as the dust worldvolume turning points stay
above the AdS curvature radius and the horizon radius; ${\bar r} > \ell, r_h$. The dust worldvolumes have little effect on the entropy, which controls 
the balance of saddle points in the microcanonical formalism. 
The reason for this is that the classical action of any  $n$-replica wormhole based on an Euclidean black-hole manifold can be analyzed as $I(X) = \beta M_{\rm eff} - A_h /4G$, where $A_h$ is the horizon area and $M_{\rm eff}$ is a generalization of the  ADM mass. The contribution $A_h /4G$ is obtained locally as coming from the  $\epsilon \to 0$ limit of the action of ${D}_\epsilon \times {\bf S}^{d-1}$, where $D_\epsilon$ is a small disc cut out an $\epsilon$ distance from the horizon. The remaining piece of the form $\beta M_{\rm eff}$, containing all the effects of the dust shells, is a renormalization of the ADM mass and does not contribute significantly to the entropy. 

For wormholes made from  $X_0$ manifolds, the main effect of the dust shells is to excise a part of the asymptotic optical volume. We can estimate this as the optical volume above radius ${\bar r}$, and contained inside a $\tau$-opening angle $\Delta\tau_W$ for each $W$. This is of order $\sum_i \ell^d \Delta \tau_i \,(\ell /{\bar r}_i)$, so the correction to $I_{\rm rad} (X_0)$ is only of relative order $|\Delta I / I | \sim \sum_i \ell^3 / \beta\, {\bar r}_i^2$, again making a small contribution to the entropy when the dust shells sit at high but finite radius. Finally, the logarithmic terms coming from quasi-zero modes or the conformal anomalies make contributions that are smaller than $I_{\rm rad}$ in the high-temperature regime of interest.  We thus conclude that the balance of saddle-point dominance will be very similar to that shown in \cref{fig:5}. 

The conditions of local instability are also expected to be well approximated by those of the universality limit, provided ${\bar r} > r_h, \ell$. This is because the dust shells affect
the low-lying spectrum of the fluctuation operator quite mildly, i.e. the Jeans instability only sets in on length scales smaller than $\ell$, and the GPY negative mode is well localized
in a region of size $\beta<\ell$ at the `tip of the cigar'.

We are now ready to take the last step, specifying the steepest-descent contour of integration for the $\beta_s$ variables in (\ref{simpleone}). Around any saddle point  $\beta(E_s)$ 
with positive heat capacity, such as the $X_0$-dominated low energy region, or the $X_+$-dominated high energy region, the integrand has no sources of complex phases from negative eigenvalues and 
has a behavior
\begin{equation}\label{integ}
\prod_{s = L, R} \Gamma \,e^{-E_s \sum_W \Delta \tau_W} e^{\beta(E_s) E_s - I(\beta(E_s))} \,e^{{1\over 2} C^{(s)}_V (1- \beta_s /\beta(E_s))^2 +\dots}\;,
\end{equation}  
where $C^{(s)}_{V} = \beta(E_s)^2 \,\partial^2_\beta\, \log {\overline Z} |_{\beta(E_s)}$  is the heat capacity with respect to variations of the inverse bulk temperature  on each side. The integrand has a local minimum at $\beta(E_s)$ and its analytic continuation into the complex $\beta_s$ plane has a path of steepest-descent passing through $\beta(E_s)$ in the direction parallel to the imaginary axis (see \cref{fig:6}),  so that the Laplace integral gives a real positive answer. 
\begin{figure}
\centering
    \begin{tikzpicture}[thick,scale=1.5,anchor=base, baseline]
    \def\xr{3}
    \def\yr{2.75}
    \draw[->,mygray, line width=0.3] (-\xr+2.5,0) -- (\xr+1.65,0) node [below ] {${\rm Re}(\beta)$};
    \draw[->,mygray, line width=0.3] (0,-\yr) -- (0,\yr) node[above left] {${\rm Im}(\beta)$};
    \draw[] (4,2.7) circle (0pt) node [above right]{${\beta_s}$};
    \draw[] (4,2.7) --(4.4,2.7){};
    \draw[] (4,2.7) --(4,3.1){};
    \draw[blue!60!black,line width=1.2,decoration={markings,mark=between positions 0.4 and 1 step 1 with \arrow{>}},postaction={decorate}] (3.9,-2.4) -- (3.9,0);
    \draw[blue!60!black,line width=1.2] (1.9,-0.25)-- (1.9,1.1)node [below right]{};
    \draw[blue!60!black,line width=1.2,decoration={markings,mark=between positions 0.32 and 1 step 1 with \arrow{<}},postaction={decorate}] (1.9,-0.25)to[out=-90,in=160] (2.1,-0.8)node [below right, scale=1.15]{$\gamma_{+}$} to[out=-20,in=230] (3.1,0.4);
    \draw[blue!60!black,line width=1.5,decoration={markings,mark=between positions 0.4 and 1 step 1 with \arrow{>}},postaction={decorate}] (3.9,-2.4)node [above right, scale=1.2]{$\gamma_{\pm}$} -- (3.9,0);
    \draw[blue!60!black,line width=1.5,decoration={markings,mark=between positions 0.01 and 1 step 0.9 with \arrow{>}},postaction={decorate}] (1.9,1.1)-- (1.9,2.4);
    \draw[blue!60!black,line width=1.2,decoration={markings,mark=between positions 0.3 and 1 step 1 with \arrow{<}},postaction={decorate}] (1.9,1.1)to[out=-90,in=70] (1,0.35)node [left, scale=1.15]{$\gamma_{-}$} to[out=-110,in=180] (1.8,-0.005) to[out=0,in=230] (3.1,0.4);
    \draw[blue!60!black,line width=1.5,decoration={markings,mark=between positions 1 and 1 step 1 with \arrow{>}},postaction={decorate}] (3.9,0)arc(0:130:0.497);
   \draw[] (3.5,0) circle (0pt) node [cross=4pt,red]{};
    \draw[red,dashed,line width=1.5] (3.5,-0.05)--(3.5,-2.6);
    \filldraw[red] (3.5,0) circle (0pt) node [above ]{${\bar\beta}$};
    \filldraw[black] (1.9,0) circle (1.8pt) node [above right ]{$ $};
    \filldraw[black] (1.9,0.05) circle (0pt) node [above right ]{$\beta(E_s)$};
    \end{tikzpicture}
 \caption{\small Contours  $\gamma_\pm$ around a saddle $\beta(E_s)$ in the complex $\beta_s$ plane, for saddles of positive $(\gamma_+)$ and negative $(\gamma_-)$ specific heat, avoiding the
 branch cut emanating from ${\bar\beta}$, the maximal inverse temperature of the black hole manifolds.} 
 \label{fig:6} 
\end{figure}

The non-trivial case occurs for the intermediate energies $E_C < E_s < E_\ell$. In this band,  the dominant wormhole manifold has a GPY negative mode supported at the tip of each $X_{\mu^-}$ core component. It contributes a factor of $1/\sqrt{\lambda^-_{\rm GPY}}$, of order $\beta_s /i$ for $\beta_s \ll \ell$. The integrand of (\ref{simpleone}) still looks like  (\ref{integ}) with two differences: the heat capacity is negative\footnote{The heat capacity is now defined as  $\beta^2 \partial^2_\beta\,\log |{\overline Z}|$, where we have removed the factor of $1/i$ from the partition function.} $C^{(s)}_V <0$, and there is the additional  GPY  factor of $1/i$. The negative heat capacity implies that the integrand has a local maximum in the real line at $\beta(E_s)$, so that
the path of steepest-descent around $\beta(E_s)$ is now tangential to the real axis.  If the contour is oriented towards the negative direction of the real axis, as shown in Figure 6,  we collect a further factor of $-1$, which combines with the GPY factor of $1/i$ and the factor of $1/i$ in the integration measure to yield again a positive definite answer. 

The detailed structure of the contours picking up the saddle points dominated by the black hole manifolds is somewhat non-trivial, since any such saddles appear `to the left' of  $\bar\beta$,  the maximum of the inverse temperature function $\beta(\mu)$, which  is actually a branch point for the analytically continued integrand $\exp(-I_\pm) /\sqrt{\lambda^\pm_{\rm GPY}}$, where the $\pm$ label refers to either of the two manifolds whose core is aproximated by $X_{\mu^\pm}$. To see this, we notice that  the GPY eigenvalue  must migrate from positive to negative values as we pass continuously  from large to small black holes by varying the mass parameter $\mu$, i.e. we have   $\lambda^\pm_{\rm GPY} \sim \pm ({\bar \beta}-\beta)/{\bar \beta}^{3}$ in the vicinity of $\bar\beta$, so that  $\sqrt{\lambda^\pm_{\rm GPY}}$ has a branch cut emanating from $\bar\beta$ for either of the two manifolds.  Perhaps more surprisingly, the classical actions  of  manifolds with black holes at the core also feature a branch cut at $\bar\beta$ when written as functions of $\beta_s$, at least in the universality limit, when they become equal to the standard  AdS black hole actions. 

This  can be argued as follows. First notice that the Euclidean action of an AdS black hole as a function of the mass parameter $I(\mu)$ satisfies 
\begin{equation}\label{chain}
{dI \over d\mu} = {dI\over d\beta} {d\beta \over d\mu} = M\,{d\beta \over d\mu}\;,
\end{equation}
where $M= (d-1) {\rm Vol}({\bf S}^{d-1} )\mu/ 16\pi G$ is the ADM mass of the black hole, always positive. Thus the functions $I(\mu)$ and $\beta(\mu)$ share the same monotonicity properties and have a maximum at the same value of $\mu$. 
This means that, on solving for  
the parameter $\mu$ to define the actions of large $(+)$ and small $(-)$ black holes as function of the inverse temperature, $I_\pm (\beta)$, we obtain the following behavior near the maximal inverse temperature $\bar \beta$: 
\begin{equation}\label{bc}
I_\pm (\beta) = {\bar I}_\pm - a \left({{\bar\beta}-\beta \over \ell}\right) \pm b \left({{\bar\beta}-\beta \over \ell}\right)^{3/2} + \dots \;,
\end{equation}
with $a$ and $b$ some positive constants of $O(1)$. We thus see that the analytically continued actions $I_\pm (\beta)$  feature a  double-valued branch cut
emerging from  $\bar\beta$. When  crossing the cut into the second sheet, the two actions, corresponding to small and large black holes, are exchanged with one another. 

The upshot of this discussion is that the branch point at $\bar\beta$, as well as the cut emerging from it, must be avoided when deforming the integration contours in search
for the saddle points. A convenient way of doing so is depicted in Figure 6, by placing the cut in the direction of the negative imaginary axis. With these provisos, the final semiclassical evaluation of (\ref{simpleone})  gives  
\begin{equation} \label{finse}
\overline {\prod_{\rm cyclic} \langle {\tilde \psi}_{i_j} | {\tilde \psi}_{i_{j+1}} \rangle}\sim  \prod_{s=L, R}  {\Gamma \beta(E_s) \over \sqrt{2\pi |C^{(s)}_V|}} \,e^{-E_s \sum_W \Delta \tau_W}
\,e^{\beta(E_s) E_s - I(\beta(E_s))}\;, 
\end{equation}
up to numerical factors of $O(1)$. In the universality limit, the exponential terms approach $\exp (S(E_L) + S(E_R)) \, \prod_j {\rm C}_j$. Therefore, we obtain the expected result, with the mathematical formalism handling successfully the `threat of un-welcome $i$ factors'. 
It should be noted that our treatment involves two successive saddle-point approximations, one in the GPI with canonical boundary conditions, and another one in the inverse Laplace transform that implements the microcanonical projection. It would be interesting to address these questions in the formalism developed in \cite{MarolfSantos}, which defines a direct microcanonical version of the GPI, free from the GPY instabilities from the beginning.

\section{Discussion}

In this note we have revisited the interesting twist proposed in refs \cite{BLMS} on the classic GPI evaluation of the Bekenstein--Hawking entropy formula. We have shown
that the method based on estimating overlaps of carefully chosen microstates, after a coarse-graining implicit in the GPI implementation of AdS/CFT rules, survives potential pitfalls
associated to unstable modes that could {\it a priori} threaten the straightforward interpretation of some path integrals as positive norms. We have seen that negative eigenvalues
of fluctuation operators around the wormhole solutions may be handled in the same fashion as in standard treatments of the density of states, such as \cite{HP}. In fact, the crucial role of the
factorization implied by the so-called `universality limit' ends up trivializing the answer to a large extent.

An interesting question about the universality limit is whether it can be liberated from the crutches of the AdS/CFT rules. The use of microcanonical states allows us to
zoom this method into the bands of states with unstable thermodynamics, such as small black holes in AdS, which have horizons much smaller than the `AdS box' and thus seem to achieve the feat.  In this case, there is a hierarchy of length scales in the overlap construction, namely one has $r_h \ll \ell \ll {\bar r}$, with the black hole radius $r_h$ being the smallest scale and the dust turning point ${\bar r}$ the largest one. In order to
 run this method without the AdS box, one would have to implement the hierarchy $r_h \ll {\bar r} \ll \ell$, i.e. take the flat space limit in the bulk {\it before}
the universality limit $m\to \infty$. It is easy to see that this becomes problematic by a simple examination of the time-lapse equation (\ref{dangle}). Setting $f(r) = 1-\mu/r^{d-2}$ and defining $x=r/{\bar r}$, we have 
$$
\Delta \tau = 2\, {\bar r} \int_{1}^\infty {dx \,x^{2-d} \over \sqrt{1-x^{4-2d}}}\;.
$$
This expression diverges for $d=2$. Even for $d>2$, the $\tau$-extent of the dust worldvolume is proportional to ${\bar r}$. Hence, taking ${\bar r} \to \infty$ as demanded
by the universality limit sends $\Delta \tau \to \infty$ and the dust `spirals' many times around the flat thermal cylinder. Such configurations do not have a straightforward interpretation after continuation back to Lorentzian signature.   It would be interesting to determine if other constructions exist that remove the AdS box from the exterior geometries while keeping it in the interior geometry, as crucially needed for this method to work. 

A final observation concerns the relation between the `implicit' GPI coarse-graining and the `explicit' one, consisting in the randomization of operator matrix elements. One would be tempted the assume that one is the strong 't Hooft coupling limit of the other in the context of AdS/CFT. One peculiar aspect of the operator averaging is that, while the {\it ansatz} (\ref{microcgb}) appears natural (ETH-like)  for a narrow high-energy  band, random operators look much more exotic when dialing the energy band to the low-energy sector. Admittedly, the microcanonical states whose overlaps are being computed in that case are quite exotic as well \cite{Martincosmo}. At any rate, it would be interesting to understand better the rules of GPI averaging, especially in low-energy sectors, and compare them with the results in \cite{Witten}.  

\subsubsection*{Acknowledgements}
We would like to thank discussions and advice from J. Magan and M. Sasieta.  This work is partially supported by the Severo Ochoa Program for Centers of Excellence through the grant CEX2020-001007-S and by the grants PID2021-123017NB- I00 and PID2022-137127NB-I00, funded by MCIN/AEI/10.13039/501100011033/ FEDER, UE, and `ERDF A way of making Europe'. E. Velasco-Aja also acknowledges support from the European Union’s Horizon 2020 research and innovation programme under the Marie Sklodowska-Curie grant agreement No 860881-HIDDeN.

\end{document}